\documentclass{article}
\usepackage{spconf,epsfig,cite}
\usepackage{amsmath,graphicx,subfig}
\usepackage{amssymb}
\usepackage{tikz}
\usetikzlibrary{decorations.pathreplacing}
\usetikzlibrary{chains,arrows,shapes,spy}
\usepackage{caption}
\usepackage{enumitem}
\tikzset{
	font={\fontsize{9}{11.0476pt}\selectfont}}
\usepackage{mathtools,amssymb,cuted}
\usepackage{pgfplots}
\pgfplotsset{compat=newest}
\usepackage{romannum}

\makeatletter
\renewcommand\paragraph{\@startsection{paragraph}{4}{\z@}%
            {-2.5ex\@plus -1ex \@minus -.25ex}%
            {1.25ex \@plus .25ex}%
            {\normalfont\normalsize\bfseries}}
\makeatother
\usepackage{algpseudocode}
\usepackage{algorithm}
\algnewcommand{\Parameters}[1]{%
	\State \textbf{\underline{Parameters}:}
	\Statex \hspace*{\algorithmicindent}\parbox[t]{.8\linewidth}{\raggedright #1}
}

\algnewcommand{\Input}[1]{%
	\State \textbf{\underline{Inputs}:}
	\Statex \hspace*{\algorithmicindent}\parbox[t]{.8\linewidth}{\raggedright #1}
}

\algnewcommand{\Initialization}[1]{%
	\State \textbf{\underline{Initialization}:}
	\Statex \hspace*{\algorithmicindent}\parbox[t]{.8\linewidth}{\raggedright #1}
}

\algnewcommand{\Iteration}[1]{%
	\State \textbf{\underline{Iteration}:}
	\Statex \hspace*{\algorithmicindent}\parbox[t]{.8\linewidth}{\raggedright #1}
}
\definecolor{r}{rgb}{0, 0, 0}
\newcommand{\pr}[1]{\ensuremath{\left[#1\right]}}
\newcommand{\pc}[1]{\ensuremath{\left(#1\right)}}
\newcommand{\chav}[1]{\ensuremath{\left\{#1\right\}}}
\newcommand{\PM}[1]{\ensuremath{\left|#1\right|}}

\title{Iterative Detection and Decoding for Multiuser MIMO Systems with Low Resolution Precoding and PSK Modulation}

\name{Erico~S.~P.~Lopes and Lukas~T.~N.~Landau}
\address{Centre for Telecommunications Studies\\
	Pontifical Catholic University of Rio de Janeiro, 
	Rio de Janeiro, Brazil 22453-900\\
	Email: erico;lukas.landau@cetuc.puc-rio.br}

\begin{document}
\ninept

\maketitle

\begin{abstract}
Low-resolution precoding techniques have gained considerable attention 
in the wireless communications area recently. Vital but hardly discussed in literature, discrete precoding in conjunction with channel coding is the subject of this study. Unlike prior studies, we propose three different soft detection methods and an iterative detection and decoding scheme that allow the utilization of channel coding in conjunction with low-resolution precoding. Besides an exact approach for computing the extrinsic information, we propose two approximations with reduced computational complexity. Numerical results based on PSK modulation and an LDPC block code indicate a superior performance as compared to the system design based on the common AWGN channel model in terms of bit-error-rate. 
\end{abstract}
%

\section{Introduction}
\label{sec:Introduction}
Multiple-input multiple-output (MIMO) systems are expected to be vital for the future of communications \cite{6G_Future_Directions}. However, the energy consumption and costs related to having multiple radio frequency front ends (RFFEs) present a challenge for this technology \cite{Power_consumption}. 

Energy efficiency is a key requirement for the next generation of wireless communications. According to \cite{6G_Use_cases}, 6G networks will require 10 to 100 times higher energy efficiency when compared to 5G. 
Another demand for future networks is higher data reliability \cite{6G_research_challanges}.\looseness-1

In these circumstances, a challenge for MIMO systems is lowering the energy consumption and costs related to the large number of RFFEs with minimum bit-error-rate (BER) compromise.

One approach to realize low energy and hardware related costs is the consideration of low-resolution data converters. Depending on the pathloss, the converters can be one of the most energy consuming elements of a RFFE, and, since their consumption scales exponentially with its resolution in amplitude \cite{Walden_1999}, using low-resolution might be favorable. However, the adoption of low-resolution converters can cause performance degradation in the BER. 

Thus, several discrete precoding approaches have been proposed in literature. Linear approaches, such as the phase Zero-Forcing (ZF-P) precoder \cite{ZF-Precoding}, benefit from a relatively low computational complexity. However, they yield performance degradation in BER especially for higher-order modulation \cite{M_Joham_ZF,Mezghani2009,one_bit_zf_saxena}.
More sophisticated nonlinear suboptimal approaches have been presented for the downlink (DL) system in \cite{MSM_precoder,CVX-CIO,GEMM_Shao2019, mezghani2020massive,Wang_Finite_alphabet,Magiq}.  
Furthermore, some optimal discrete precoding algorithms exist in the literature such as \cite{Landau2017,General_MMDDT_bb,Jacobsson2018,MMSE_bb_precoder}.

Both, optimal and suboptimal nonlinear approaches show promising results in terms of uncoded BER. Yet, practical systems usually employ coding schemes with soft detection to provide a higher degree of reliability. \textcolor{r}{Soft detection in conjunction with discrete precoding was first considered in \cite{jacobsson2018nonlinear}, where convolutional codes are decoded using a BCJR algorithm in the context of OFDM.}

\textcolor{r}{Different from \cite{jacobsson2018nonlinear}, which for computing the log-likelihood-ratios (LLRs) relies on the common method for AWGN channels, this study proposes three sophisticated approaches that compute extrinsic information considering the effects of the discrete precoder in the probability density function (PDF) of the received signal. The extrinsic information is then used for computing the LLRs via the discrete precoding aware (DPA) iterative detection and decoding (IDD) algorithm.}

{The first method computes the extrinsic information based on the true probability density function (PDF) of the received signal. The second, relies on a nonlinear Gaussian approximation of the original PDF for its computation. Finally, the third relies on a description of the received signal by a linear model with a Gaussian additive distortion term.}


The rest of the paper is organized as follows: Section~\ref{sec:system_model} describes the system model. Section~\ref{sec:receiver} exposes the receiver design. Section~\ref{sec:numerical_results} exposes numerical results, while Section~\ref{sec:conclusions} presents the conclusions.

\section{System Model}
\label{sec:system_model}
\begin{figure} [t]
\centering
\input{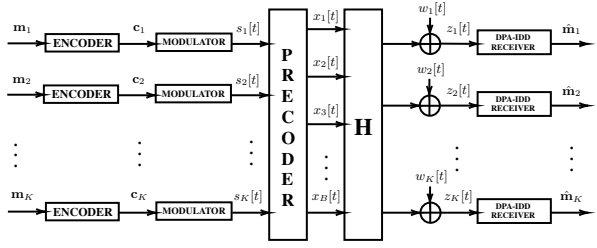}
\caption{Multiuser MIMO DL with discrete precoding and channel coding}
\label{fig:system_model}       
\end{figure}

This study considers a single-cell Multiuser MIMO DL system in which the BS has perfect channel state information (CSI) and is equipped with $B$ transmitting antennas which serves $K$ single antenna users as  illustrated by Fig.~\ref{fig:system_model}.

A blockwise transmission is considered in which the BS delivers $N_b$ bits for each user. The user specific block is denoted by the vector $\boldsymbol{m}_k=[m_{k,1} \hdots m_{k,N_b}]$, where the index $k$ indicates the $k$-th user. Each vector $\boldsymbol{m}_k$ is encoded into a codeword vector denoted by $\boldsymbol{c}_k=[{c}_{k,1} \hdots {c}_{k,\frac{N_b}{R}}]$, where $R$ is the code rate.{ A systematic encoding operation is considered meaning that 
$\boldsymbol{c}_k=[{p}_{k,1}, \hdots, {p}_{k,\frac{N_b(1-R)}{R}}, m_{k,1}, \hdots m_{k,N_b}]$, where ${p}_{k,i}$ is the $i$-th parity bit.}\looseness-1
 
Each encoder provides, sequentially over time slots, $M$ bits to a modulator which maps them into a symbol $s\in \mathcal{S}$ using Gray coding. The set $\mathcal{S}$ represents all possible symbols of a $\alpha_{s}$-PSK modulation and is described by 
\begin{align}
	    \mathcal{S}= \left\{s: s= \text{e}^  \frac{j\pi (2 i+1) }{\alpha_{s}}  \textrm{,  for  }  i=1,\ldots, \alpha_{s} \right\}  \textrm{,}
	    \label{S_set}
	\end{align}
where $\alpha_s=2^M$. The mapping operation is denoted as $s[t]=\mathcal{M}\pc{\boldsymbol{r}_{k,t}}$, where $\boldsymbol{r}_{k,t}=\pr{r_{k,t,1}, \hdots, {r}_{k,t,M}}$ is the $t$-th bit vector, taken from $\boldsymbol{c}_k$. The vector $\boldsymbol{r}_{k,t}$ can also be expressed as $\boldsymbol{r}_{k,t}=\pr{c_{k,(t-1)M+1}, \hdots, {c}_{k,t M}}$ for $t=1,\hdots, \frac{N_b}{RM}$.
After mapping, the symbols of all $K$ users are represented in a stacked vector notation as $\boldsymbol{s}[t]=\pr{s_{1}[t] \hdots s_{K}[t]}^{T} \in \mathcal{S}^K $ for each time slot $t$.

The vector $\boldsymbol{s}[t]$ is forwarded to the precoder, which computes the transmit vector $\boldsymbol{x}[t]=[x_{1}[t] \hdots x_{B}[t]]^{T} \in \mathcal{X}^{B}$. The entries of the transmit vector are constrained to the set $\mathcal{X}$, which describes an $\alpha_{x}$-PSK alphabet, denoted by
\begin{align}
	    \mathcal{X}= \left\{x: x= \text{e}^  \frac{j\pi (2 i+1) }{\alpha_{x}}  \textrm{,  for  }  i=1,\ldots, \alpha_{x} \right\} \textrm{.}
	    \label{X_set}
\end{align}

A flat fading channel described by the matrix $\boldsymbol{H}$ with coefficients $h_{k,b}$ is considered, where $k$ and $b$ denote the index of the user and the transmit antenna, respectively.
A block fading model is considered in which $\boldsymbol{H}$ is invariant during the transmission time.

The BS computes for each coherence time interval of the channel the lookup-table $\mathcal{L}$ containing all possible precoding vectors, which then implies $\boldsymbol{s} \in \mathcal{S}^K \iff \boldsymbol{x}(\boldsymbol{s}) \in \mathcal{L}$.


At the user terminals the received signals are distorted by additive white Gaussian noise (AWGN) denoted by the complex random variable ${w_k}\pr{t}\sim \mathcal{CN}  ({0},\sigma_w^2)$. The received signal from the $k$-th user is denoted by $
    {z_k}\pr{t} = \boldsymbol{h}_k\ \boldsymbol{x}\pr{t}+{w_k}\pr{t}\text{,}
$
where $\boldsymbol{h}_k$ is the $k$-th row of the channel matrix $\boldsymbol{H}$.

Each received signal ${z}_k[t]$ is forwarded to the IDD receiver where the transmitted block will be estimated. Finally the data block available to the $k$-th user reads as $\hat{\boldsymbol{m}}_k=\pr{\hat{m}_{k,1}, \hdots, \hat{m}_{k,N_b} }$.



\section{Receiver Design}
\label{sec:receiver}
This section exposes the design of the DPA-IDD Receiver where three soft detection methods for the computation of the extrinsic information are proposed. The objective is to enable channel coding in conjunction with discrete precoding methods.

The main objective of the DPA-IDD receiver is to estimate $\hat{\boldsymbol{m}}$ by computing LLRs and performing a decision. 
In general the LLRs are defined as follows 
\begin{equation}
\label{eq:L1}
L({c}_{k,i})=\ln\pc{\frac{
{P}\pc{{c}_{k,i} = 0 | {z}_k[t]}}
{{P}\pc{{c}_{k,i} = 1 | {z}_k[t]}}}\text{,}
\end{equation}
where ${z}_k[t]$ is the received signal and  ${c}_{k,i} \in \{0,1\}$. Using Bayes' theorem, equation \eqref{eq:L1} is rewritten as
\begin{align}
\label{eq:a_posteriori_LLR}
L({c}_{k,i})&=\ln\pc{\frac{
{p}\pc{{z}_k[t]|{c}_{k,i} = 0}}
{{p}\pc{{z}_k[t]|{c}_{k,i} = 1}}
}+
\ln\pc{\frac{
{P}\pc{{c}_{k,i} = 0}}
{{P}\pc{{c}_{k,i} = 1}}
}\notag\\[5pt]
&=L_e \pc{c_{k,i}} + L_a\pc{{c}_{k,i}} ,
\end{align}
where $L_e \pc{c_{k,i}}$ and $L_a \pc{c_{k,i}}$ denote the extrinsic and a priori information functions respectively. 

\subsection{Extrinsic Information Computation}
\label{subsec:extrinsic_llr}

In this section, three methods for computing $L_e(c_{k,i})$ are presented. The first method computes the extrinsic information based on the true PDF of the received signal, while the second relies on a nonlinear Gaussian approximation of the original PDF. The third approach calculates $L_e(c_{k,i})$ by relying on a linear model. 
As shown in equation \eqref{eq:a_posteriori_LLR}, the $L_e(c_{k,i})$ is defined as 
\begin{align}
\label{eq:extrinsic_LLR_definition}
    L_e \pc{c_{k,i}}=\ln\pc{\frac{
{p}\pc{{z}_k[t]|{c}_{k,i} = 0}}
{{p}\pc{{z}_k[t]|{c}_{k,i} = 1}}
}.
\end{align}
Using the law of total probability equation \eqref{eq:extrinsic_LLR_definition} can be expanded as
\begin{align}
\label{eq:extrinsic_LLR}
L_e \pc{c_{k,i}}&=
\ln\pc{\frac{
\displaystyle \sum_{s\in S_0}{p}\pc{{z}_k[t]|s} {P}\pc{s|{r}_{k,t,\upsilon} = 0}}
{\displaystyle \sum_{s\in S_1}{p}\pc{{z}_k[t]|s}{P}\pc{s|{r}_{k,t,\upsilon} = 1}}
}\text{,}
\end{align}
with index $\upsilon={i-(t-1)M} \in \{1,\hdots, M\}$. The sets $S_{0}$ and $S_{1}$ represent all possible constellation points where the $\upsilon$-th bit of $\boldsymbol{r}_{k,t}$ is 0 or 1, respectively.
For a given $s\in S_g$, $g \in \{0,1\}$, if $\mathcal{M}^{-1}(s)=[a_{1},\hdots, a_{\upsilon}=g, \hdots,a_{M}]$, the probability ${P}\pc{s|{r}_{k,t,\upsilon}=g}$ reads as
\begin{align}
    \label{eq:a_priori_llr}
    {P}\pc{s|{r}_{k,t,\upsilon}=g}=\displaystyle\prod_{\substack{l=1 \\[2pt] l\neq \upsilon}}^{M} \frac{e^{-2\pc{a_l-\frac{1}{2}}L_a({r}_{k,t,l})}}
    {1+e^{-2\pc{a_l-\frac{1}{2}}L_a({r}_{k,t,l})}} ,
\end{align}
where $L_a\pc{{r}_{k,t,l}}=L_a\pc{{c}_{k,l+(t-1)M}}$. Note that for computing $L_e(c_{k,i})$, the channel law $p\pc{{z}_k[t]|s}$, for all $s\in \mathcal{S}$, and the a priori information function $L_a\pc{c_{k,i}}$, for $i=1,\hdots, \frac{N_b}{R}$, are required.

\subsubsection{Discrete Precoding Aware Soft Detector}
\label{subsec:optimal_llr}
In this subsection, we introduce the DPA Soft Detector as a method for computing $L_e(c_{k,i})$, which can be considered as the soft MAP detector. 
First, the received signal $z_k[t]$ is rewritten in a stacked vector notation 
$\boldsymbol{z}_r[t]= [ \text{Re}\chav{z_k[t]} \  \text{Im}\chav{z_k[t]} ]^T$, where, for simplicity, the index $k$ is suppressed. 
The distribution $p\pc{{z}_k[t]|s}$ is given by
\begin{align}
    \label{eq:real_pdf}
    p\pc{{z}_k[t]|s}&=\displaystyle\sum_{\boldsymbol{s^\prime}\in \mathcal{S}^{K-1}} p\pc{{z}_k[t]|s,\boldsymbol{s}^\prime} P(\boldsymbol{s}^\prime)\\
    &=\pc{\frac{1}{\alpha_s}}^{K-1} \frac{1}{\pi \sigma_w^2}\displaystyle\sum_{\boldsymbol{s^\prime}\in \mathcal{S}^{K-1}}
    \text{e}^{-\frac{\PM{\PM{\boldsymbol{z}_r[t]-\text{E}\chav{\boldsymbol{z}_r[t]|s,\boldsymbol{s}^\prime}}}_2^2}{\sigma_w^2}} \notag,
\end{align}
where $\boldsymbol{s}^\prime=\pr{s_1^\prime,\hdots,s_{k-1}^\prime,s_{k+1}^\prime,\hdots,s^\prime_K}^T$ corresponds to the symbols of the other users. For a given $s$ and $\boldsymbol{s}^\prime$ the expected value of the receive signal is given by
\begin{align}
    \boldsymbol{\mu}_{\boldsymbol{z}_r|\boldsymbol{s}}=\text{E}\chav{\boldsymbol{z}_r[t]|s,\boldsymbol{s}^\prime}=[\text{Re}\chav{\boldsymbol{h}_k\boldsymbol{x}\pc{s,\boldsymbol{s}^\prime}} \ \ \text{Im}\chav{\boldsymbol{h}_k\boldsymbol{x}\pc{s,\boldsymbol{s}^\prime}}]^T.\notag
\end{align}
With this, $L_e(c_{k,i})$ can be computed by inserting \eqref{eq:real_pdf} into \eqref{eq:extrinsic_LLR}. The resulting expression, finally, reads as
\begin{align}
\label{eq:optimal_llr}
    L_e(c_{k,i})&=\ln\pc{\frac{
\displaystyle \sum_{s\in S_{0}}
\displaystyle\sum_{\boldsymbol{s^\prime}\in \mathcal{S}^{K-1}}
    \hspace{-1em}\text{e}^{-\frac{\PM{\PM{\boldsymbol{z}_r[t]-\boldsymbol{\mu}_{\boldsymbol{z}_r|\boldsymbol{s}}}}_2^2}{\sigma_w^2}}
{P}\pc{s|{r}_{k,t,\upsilon} = 0}}
{\displaystyle \sum_{s\in S_{1}}
\displaystyle\sum_{\boldsymbol{s^\prime}\in \mathcal{S}^{K-1}}
    \hspace{-1em}\text{e}^{-\frac{\PM{\PM{\boldsymbol{z}_r[t]-\boldsymbol{\mu}_{\boldsymbol{z}_r|\boldsymbol{s}}}}_2^2}{\sigma_w^2}}
{P}\pc{s|{r}_{k,t,\upsilon} = 1}}
}.
\end{align}
Note that, for using \eqref{eq:optimal_llr}, $p\pc{{z}_k[t]|s,\boldsymbol{s}^\prime}$ is evaluated for all members of $\mathcal{S}^{K-1}$. Hence, computing \eqref{eq:optimal_llr} can lead to a prohibitive computational complexity at the receiver side for systems with many users. 

\subsubsection{{Gaussian Discrete Precoding Aware Soft Detector}}
\label{subsec:exact_llr}
{To reduce the computational complexity we introduce the Gaussian Discrete Precoding Aware (GDPA) Soft Detector.} 
The basic assumption is that the vector $\boldsymbol{z}_r[t]$ can be described as a Gaussian random vector, meaning
\begin{align}
\label{eq:channel_probability}
    \tilde{p}\pc{{z}_k[t] |   s}=\frac{ \text{e}^{-\frac{1}{2} 
   \pr{\pc{\boldsymbol{z}_r[t]-\tilde{\boldsymbol{\mu}}_{z_r|s}} ^T\boldsymbol{C}_{z_r|s}^{-1}\pc{\boldsymbol{z}_r[t]-\tilde{\boldsymbol{\mu}}_{z_r|s}}}} }{2\pi \sqrt{\text{det}\pc{\boldsymbol{C}_{z_r|s}}}}.
\end{align}
In the following the computation of $\tilde{\boldsymbol{\mu}}_{z_r|s}$ and $\boldsymbol{C}_{z_r|s}$ is detailed. Since $\text{E}\chav{\text{Re}\chav{a}}=\text{Re}\chav{\text{E}\chav{a}}$, and $\text{E}\chav{\text{Im}\chav{a}}=\text{Im}\chav{\text{E}\chav{a}}$, we compute first the expected value of the complex received signal, which reads as
\begin{align}
\label{eq:mean}
\text{E}\chav{z_k[t]|  s}
&=\text{E}\chav{{ {\boldsymbol{h}_k\boldsymbol{x}[t]}}\ |  s}.
\end{align}
In order to simplify the notation we introduce the variable $\zeta(\boldsymbol{s})=\boldsymbol{h}_k\ \boldsymbol{x}\pc{\boldsymbol{s}}$. The mean vector $\tilde{\boldsymbol{\mu}}_{z_r|s}$ is, then, given by
\begin{align}
\label{eq:mean_vector}
\tilde{\boldsymbol{\mu}}_{z_r|s}&=\begin{bmatrix} 
    \text{E}\chav{\text{Re}\chav{\zeta\pc{\boldsymbol{s}}} |  s}  \ \ \ 
    \text{E}\chav{\text{Im}\chav{\zeta\pc{\boldsymbol{s}}} |  s}
\end{bmatrix} ^T,
\end{align}
where
\begin{align}
\label{eq:expected_values1}
\text{E}\chav{\text{Re}\chav{\zeta\pc{\boldsymbol{s}}}|{s}}&=\pc{\frac{1}{\alpha_s}}^{K-1}\displaystyle \sum_{\boldsymbol{s}\in \mathcal{D}}{\text{Re}\chav{\zeta\pc{\boldsymbol{s}}}},\\
\label{eq:expected_values2}
\text{E}\chav{\text{Im}\chav{\zeta\pc{\boldsymbol{s}}}|{s}}&=\pc{\frac{1}{\alpha_s}}^{K-1}\displaystyle \sum_{\boldsymbol{s}\in \mathcal{D}}{\text{Im}\chav{\zeta\pc{\boldsymbol{s}}}}
\end{align}
and $\mathcal{D}$ is the set of all possible $\boldsymbol{s}[t]$ whose k-th entry is $s$. Moreover, the corresponding covariance matrix is given by
\begin{align}
\label{eq:czr|s}
    \boldsymbol{C}_{z_r|s}=\begin{bmatrix} 
    \sigma_{r|s}^2\ \ \ &\rho_{ri|s}  \\  
    \rho_{ri|s} \ \ \ &\sigma_{i|s}^2 
\end{bmatrix}.
\end{align}
The entries of $\boldsymbol{C}_{z_r|s}$ read as
 \begin{align}
      \sigma_{r|s}^2&=\frac{\sigma_w^2}{2}+ \text{E}\chav{\text{Re}\chav{\zeta\pc{\boldsymbol{s}}}^2|{s}}- \text{E}\chav{\text{Re}\chav{\zeta\pc{\boldsymbol{s}}}|{s}}^2,  \\[5pt]
      \sigma_{i|s}^2&=\frac{\sigma_w^2}{2}+\text{E}\chav{\text{Im}\chav{\zeta\pc{\boldsymbol{s}}}^2|{s}}- \text{E}\chav{\text{Im}\chav{\zeta\pc{\boldsymbol{s}}}|{s}}^2, \\[5pt]
     \rho_{ri|s}&=\text{E}\{\text{Re}\{\zeta\pc{\boldsymbol{s}}\}\text{Im}\{\zeta\pc{\boldsymbol{s}}|s\}\}-\notag\\&\hspace{5em}\text{E}\{\text{Re}\{\zeta\pc{\boldsymbol{s}}\}|s\}\text{E}\{\text{Im}\chav{\zeta\pc{\boldsymbol{s}}}|s\},
 \end{align}
where 
\begin{align}
\text{E}&\chav{\text{Re}\chav{\zeta\pc{\boldsymbol{s}}}^2|{s}}=\pc{\frac{1}{\alpha_s}}^{K-1}\displaystyle \sum_{\boldsymbol{s}\in \mathcal{D}}{\text{Re}\chav{\zeta\pc{\boldsymbol{s}}}^2},\notag\\
\text{E}&\chav{\text{Im}\chav{\zeta\pc{\boldsymbol{s}}}^2|{s}}=\pc{\frac{1}{\alpha_s}}^{K-1}\displaystyle \sum_{\boldsymbol{s}\in \mathcal{D}}{\text{Im}\chav{\zeta\pc{\boldsymbol{s}}}^2},\notag\\
\text{E}&\chav{\text{Re}\chav{\zeta\pc{\boldsymbol{s}}}\text{Im}\chav{\zeta\pc{\boldsymbol{s}}}|s}=
{\frac{1}{\alpha_s^{K-1}}}\displaystyle \sum_{\boldsymbol{s}\in \mathcal{D}}{\text{Re}\chav{\zeta\pc{\boldsymbol{s}}}\text{Im}\chav{\zeta\pc{\boldsymbol{s}}}}\notag
\end{align} 
and $\text{E}\chav{\text{Re}\chav{\zeta\pc{\boldsymbol{s}}}|s}$ and $\text{E}\chav{\text{Im}\chav{\zeta\pc{\boldsymbol{s}}}|s}$ are defined in equations \eqref{eq:expected_values1} and \eqref{eq:expected_values2}, respectively. Based on $\boldsymbol{C}_{z_r|s}$ and $\tilde{\boldsymbol{\mu}}_{z_r|s}$,  $L_e(c_{k,i})$ is computed as
\begin{align}
\label{eq:LLR_GDPA}
    L_e(c_{k,i})&=\ln\pc{\frac{
\displaystyle \sum_{s\in S_{0}}
\frac{\text{e}^{\Psi_s} }
{\sqrt{\text{det}\pc{{\boldsymbol{C}_{z_r|s}}}}}
{P}\pc{s|{r}_{k,t,\upsilon} = 0}}
{\displaystyle \sum_{s\in S_{1}}
\frac{\text{e}^{\Psi_s} }
{\sqrt{\text{det}\pc{{\boldsymbol{C}_{z_r|s}}}}}
{P}\pc{s|{r}_{k,t,\upsilon} = 1}}
}\text{,}
\end{align}
where
\begin{align}
    \Psi_{s}=-\frac{1}{2} 
   \pr{\pc{\boldsymbol{z}_r[t]-\tilde{\boldsymbol{\mu}}_{z_r|s} }^T\boldsymbol{C}_{z_r|s}^{-1}\pc{\boldsymbol{z}_r[t]-\ \tilde{\boldsymbol{\mu}}_{z_r|s}}}
\end{align}
and ${P}\pc{s|{r}_{k,t,\upsilon} = g}$ for $g\in\{0,1\}$ can be computed with equation \eqref{eq:a_priori_llr} considering $L_a(c_{k,i})$ for $i=1,\hdots,\frac{N_b}{R}$.

{Note that, when calculating $L_e(c_{k,i})$ using \eqref{eq:LLR_GDPA}, $\tilde{p}({z_k}[t]|s)$ is evaluated only $\alpha_s$ times. This results in a significant decrease in computational complexity, when compared with the approach proposed in \eqref{eq:optimal_llr}. 
However, for computing \eqref{eq:LLR_GDPA}, the receiver requires access to $\boldsymbol{C}_{z_r|s}$ and $\tilde{\boldsymbol{\mu}}_{z_r|s}$ for all values of $s$. These parameters need to be provided by the BS which causes communication overhead. In this context, an alternative method that requires a fewer number of parameters to be transmitted is desired.}
\subsubsection{{Linear Model Based Discrete Precoding Aware Soft Detector}}
\label{subsec:Linear_Model}
In this subsection, a method for computing $L_e(c_{k,i})$ with a reduced number of model parameters is devised. This proposed approach relies on the description of the received signal by a linear model.\looseness-1
\paragraph{Discrete Precoding Aware Linear Model}
\label{subsec:introduction_on_weighting}
The Discrete Precoding Aware Linear Model (DPA-LM) is based on the assumption that the received signal can be expressed by 
\begin{align}
    \label{eq:channel_model}
        {z}_k[t]=h_{k}^\text{eff}{s_k[t]+w_k[t]+\epsilon_k[t]} \text{,}
\end{align}
where $h^\text{eff}_k \in \mathcal{C}$ is a factor that expresses the precoder and channel effects on the transmit symbol of the $k$-th user and $\epsilon_k[t]$ is the error term that denotes the difference between $z_k[t]$ and $h_{k}^\text{eff}s_k[t]+w_k[t]$. To identify an appropriate $h^\text{eff}_k$ we consider the following MSE optimization problem \looseness-1
 \begin{align}
 \label{eq:problem_for_heff}
     h^\text{eff}_k&=\arg\min \lambda_{\epsilon_k}^2 =\arg\min \text{E}\chav{ \PM{ \epsilon_k[t]    }^2}\notag\\
     &= \arg\min_{\gamma \in \mathcal{C  }}\text{E}\chav{ \PM{ \boldsymbol{h}_k \ \boldsymbol{x}[t] - {\gamma \ s_k[t]}}^2}\text{,}
 \end{align}
where the optimal solution is given by
 \begin{align}
    \hspace{-1.8em}\quad h^\text{eff}_k={\frac{1}{\alpha_s^K\ \sigma_s^2}}\sum_{\boldsymbol{s} \in \mathcal{S}^K} \ s_k^*(\boldsymbol{s}) \ \zeta(\boldsymbol{s}),
    \quad \lambda_{\epsilon_k}^2=
    \boldsymbol{h}_k\ \boldsymbol{\Lambda}_x \ \boldsymbol{h}^H_k-\PM{h^\text{eff}_k}^2 \sigma_s^2, \notag
 \end{align}
with $\boldsymbol{\Lambda}_x=\pc{\frac{1}{\alpha_s}}^K \displaystyle \sum_{\boldsymbol{s}\in \mathcal{S}^K} {\boldsymbol{x}\pc{\boldsymbol{s}} \boldsymbol{x}\pc{\boldsymbol{s}}^H}$ and $s_k\pc{\boldsymbol{s}}$ being the $k$-th element of $\boldsymbol{s}$.

\paragraph{DPA-LM Soft Detector}
\label{subsec:LLR_approximation}
This subsection proposes the DPA-LM Soft Detector as a method for computing the extrinsic information based on the linear model previously presented.
The strategy relies on the assumption that the error term $\epsilon_k[t]$ is a circular symmetric complex Gaussian random variable. 
The expected value of the received signal is calculated as
\begin{align}
\text{E}\chav{z_k[t]|s}&= h_k^\text{eff} s +\text{E}\chav{\epsilon_k[t]|s}, 
\end{align}
and assuming $\text{E}\chav{\epsilon_k[t]|s}=0 \ \forall \ s \in \mathcal{S}$ yields
\begin{align}    
\label{eq:muzrs}
\boldsymbol{\mu}_{z_r|s}^{\text{eff}}&=\begin{bmatrix} 
    \text{Re}\chav{h_k^\text{eff} s}  \ \ \  
    \text{Im}\chav{h_k^\text{eff} s}
\end{bmatrix}^T . 
\end{align}
Considering that $\boldsymbol{C}_{z_r}^{\text{eff}}=\frac{{\sigma}_{\text{eff}_k}^2}{2} \boldsymbol{I}$
with ${\sigma}_{\text{eff}_k}^2={\lambda}_{\epsilon_k}^2+ \sigma_w^2$ being the effective noise variance, the extrinsic information function from \eqref{eq:extrinsic_LLR} simplifies to 
\begin{align}
\label{eq:linear_model_LLR}
   \hspace{-1.885mm}
   L_e\pc{c_{k,i}}=
    \text{ln}\pc{
    \frac{\displaystyle \sum_{s\in S_{0}}  \text{e}^{-\frac{\PM{{z_k[t]-h_k^\text{eff}\ s}}^2}{{\sigma}_{\text{eff}_k}^2} } \ {P}\pc{s|{r}_{k,t,\upsilon} = 0}
}
    {\displaystyle\sum_{s\in S_{1}}  \text{e}^{-\frac{\PM{{z_k[t]-h_k^\text{eff}\ s}}^2}{{\sigma}_{\text{eff}_k}^2}}\ {P}\pc{s|{r}_{k,t,\upsilon} = 1}}
    } ,
\end{align}
where ${P}\pc{s|{r}_{k,t,\upsilon} = g}$, for $g \in \{0,1\}$ is computed via equation \eqref{eq:a_priori_llr}. \looseness-1

The computation of $L_e(c_{k,i})$ according to \eqref{eq:linear_model_LLR} only requires knowledge about the parameters $h_k^\text{eff}$ and ${\sigma}_{\text{eff}_k}^2$, which are independent of the data symbol $s$. In comparison with the method from subsection~\ref{subsec:exact_llr}, the number of parameters that need to be transmitted in advance to the information data is significantly reduced.

\subsection{DPA-IDD Scheme}
\label{subsec:dpa_idd_algorithm}

{Subsections \ref{subsec:optimal_llr}, \ref{subsec:exact_llr} and \ref{subsec:LLR_approximation} expose different methods for computing $L_e(c_{k,i})$ when $L_a(c_{k,i})$ is known.} Using these results, the DPA-IDD scheme is presented as a way of computing $L(c_{k,i})$ via making an iterative estimation of $L_a(c_{k,i})$ and, consequently, $L_e(c_{k,i})$. {For description of the DPA-IDD scheme, we define $\boldsymbol{L}=[L(c_{k,1}) \hdots L(c_{k,\frac{N_b}{R}})]$, $\boldsymbol{L}_e=[L_e(c_{k,1}) \hdots L_e(c_{k,\frac{N_b}{R}})]$ and $\boldsymbol{L}_a=[L_a(c_{k,1}) \hdots L_a(c_{k,\frac{N_b}{R}})]$.}

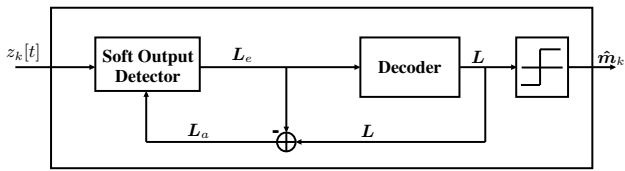
\begin{figure}[ht]
\begin{center}
\tikzset{every picture/.style={line width=0.75pt}} 

\begin{tikzpicture}[x=0.25pt,y=0.25pt,yscale=-1,xscale=1]

\draw   (630,189) -- (779.93,189) -- (779.93,279) -- (630,279) -- cycle ;
\draw    (108,234) -- (226,234) ;
\draw [shift={(229,234)}, rotate = 180] [fill={rgb, 255:red, 0; green, 0; blue, 0 }  ][line width=0.08]  [draw opacity=0] (8.93,-4.29) -- (0,0) -- (8.93,4.29) -- cycle    ;
\draw   (162,144) -- (981,144) -- (981,387) -- (162,387) -- cycle ;
\draw    (819,234) -- (819,347.23) ;
\draw    (535,347.23) -- (819,347.23) ;
\draw [shift={(532,347.23)}, rotate = 0] [fill={rgb, 255:red, 0; green, 0; blue, 0 }  ][line width=0.08]  [draw opacity=0] (8.93,-4.29) -- (0,0) -- (8.93,4.29) -- cycle    ;
\draw   (229,189) -- (387,189) -- (387,270) -- (229,270) -- cycle ;
\draw    (781,234) -- (864,234) ;
\draw [shift={(867,234)}, rotate = 180] [fill={rgb, 255:red, 0; green, 0; blue, 0 }  ][line width=0.08]  [draw opacity=0] (8.93,-4.29) -- (0,0) -- (8.93,4.29) -- cycle    ;
\draw    (387,234) -- (627,234) ;
\draw [shift={(630,234)}, rotate = 180] [fill={rgb, 255:red, 0; green, 0; blue, 0 }  ][line width=0.08]  [draw opacity=0] (8.93,-4.29) -- (0,0) -- (8.93,4.29) -- cycle    ;
\draw    (306,273) -- (306,347.23) ;
\draw [shift={(306,270)}, rotate = 90] [fill={rgb, 255:red, 0; green, 0; blue, 0 }  ][line width=0.08]  [draw opacity=0] (8.93,-4.29) -- (0,0) -- (8.93,4.29) -- cycle    ;
\draw    (306,347.23) -- (504,347.23) ;
\draw   (865.8,189) -- (944.24,189) -- (944.24,279) -- (865.8,279) -- cycle ;
\draw    (874.03,234.53) -- (936,234.53) ;
\draw    (901.92,207) -- (932.9,207) ;
\draw    (873,261) -- (901.92,261) ;
\draw    (901.92,207.13) -- (901.92,261) ;
\draw    (945,234) -- (1014,234) ;
\draw [shift={(1017,234)}, rotate = 180] [fill={rgb, 255:red, 0; green, 0; blue, 0 }  ][line width=0.08]  [draw opacity=0] (8.93,-4.29) -- (0,0) -- (8.93,4.29) -- cycle    ;
\draw   (504,347.23) .. controls (504,339.37) and (510.27,333) .. (518,333) .. controls (525.73,333) and (532,339.37) .. (532,347.23) .. controls (532,355.09) and (525.73,361.46) .. (518,361.46) .. controls (510.27,361.46) and (504,355.09) .. (504,347.23) -- cycle ; \draw   (504,347.23) -- (532,347.23) ; \draw   (518,333) -- (518,361.46) ;
\draw    (518,233.77) -- (518,330) ;
\draw [shift={(518,333)}, rotate = 270] [fill={rgb, 255:red, 0; green, 0; blue, 0 }  ][line width=0.08]  [draw opacity=0] (8.93,-4.29) -- (0,0) -- (8.93,4.29) -- cycle    ;

\draw (704.97,233.32) node  [scale=0.75] [align=center] {\textbf{${\text{Decoder}}$}};
\draw (120,212.5) node  [scale=0.75]  {$z_{k}[t]$};
\draw (808,218.5) node  [scale=0.75]  {$\boldsymbol{L}$};
\draw (641,330.5) node  [scale=0.75]  {$\boldsymbol{L}$};
\draw (308.64,232.21) node  [scale=0.75] [align=center] {\textbf{Soft Output}\\\textbf{Detector}};
\draw (385.5,330.5) node  [scale=0.75]  {$\boldsymbol{L}_a$};
\draw (451.5,215.5) node  [scale=0.75]  {$\boldsymbol{L}_e$};
\draw (1010,219.5) node  [scale=0.75]  {$\boldsymbol{\hat{m}}_k$};
\draw (502,335) node  [scale=1]  {\textbf{-}};

\end{tikzpicture}
\caption{DPA-IDD Receiver Topology} 
\label{fig:dpa_idd_topology}       
\end{center}
\vspace{-1em}
\end{figure}

The principle of the proposed receiver is based on equation \eqref{eq:a_posteriori_LLR}. Based on $\boldsymbol{L}$ and $\boldsymbol{L}_e$, the a priori information is extracted via $\boldsymbol{L}_a=\boldsymbol{L}-\boldsymbol{L}_e$. 
With this, for initialization, the detector calculates $\boldsymbol{L}_e$ assuming $\boldsymbol{L}_a=\boldsymbol{0}$ and forwards it to the decoder. The decoder outputs the LLR vector $\boldsymbol{L}$. Using $\boldsymbol{L}$ and $\boldsymbol{L}_e$, the a priori information is calculated and fed back into the detector which will, then, recompute $\boldsymbol{L}_e$ based on the updated $\boldsymbol{L}_a$. This process is done recursively until the maximum number of iterations is reached. An illustration of the receiving process is shown in Fig.~\ref{fig:dpa_idd_topology}.
The DPA-IDD technique does not require a specific method for computing $\boldsymbol{L}_e$. Hence, the approaches presented in subsections \ref{subsec:optimal_llr}, \ref{subsec:exact_llr} and \ref{subsec:LLR_approximation} are compatible with the framework and can be used for calculating $\boldsymbol{L}_e$.



\section{Numerical Results}
\label{sec:numerical_results}

In this section, the proposed soft detection schemes are evaluated considering as the MMSE branch-and-bound approach from \cite{MMSE_bb_precoder} as the precoding technique.
The shown results were computed using an LDPC block code with a block size of $\frac{N_b}{R}=486$ bits and code rate $R=1/2$. The LLRs are processed by sum-product algorithm (SPA) decoders \cite{SPA-Decoder}. The examined system has $K=2$ users and $B=6$ BS antennas where the data symbols are considered as 8-PSK and the precoded symbols are considered as QPSK, meaning $\alpha_s=8$ and $\alpha_x=4$. 
We evaluate the soft detection methods in conjunction with the proposed DPA-IDD scheme. In such circumstances, the proposed soft detectors are compared with the conventional AWGN detector design, described by
\begin{align}
\label{eq:extrinsic_awgn}
   L_e\pc{c_{k,i}}=
    \text{ln}\pc{
    \frac{\displaystyle \sum_{s\in S_0}  \text{e}^{-\frac{\PM{{z_k[t]-s}}^2}{{\sigma}_{w}^2}} \ {P}\pc{s|{r}_{k,t,\upsilon} = 0}
}
    {\displaystyle\sum_{s\in S_1}  \text{e}^{-\frac{\PM{{z_k[t]-s}}^2}{{\sigma}_{w}^2}}\ {P}\pc{s|{r}_{k,t,\upsilon} = 1}}
    },
\end{align}
\textcolor{r}{which is analog to as considered in \cite{jacobsson2018nonlinear} in the context of convolutional codes with a BCJR decoder.}
The examined approaches are 
    1. Uncoded transmission;
    2. Coded transmission using the DPA soft detector \eqref{eq:optimal_llr};
    3. Coded transmission using the GDPA soft detector \eqref{eq:LLR_GDPA};
    4. Coded transmission using DPA-LM soft detector \eqref{eq:linear_model_LLR};
    5. Coded transmission using AWGN method \eqref{eq:extrinsic_awgn}.

\begin{figure}[ht]
\input{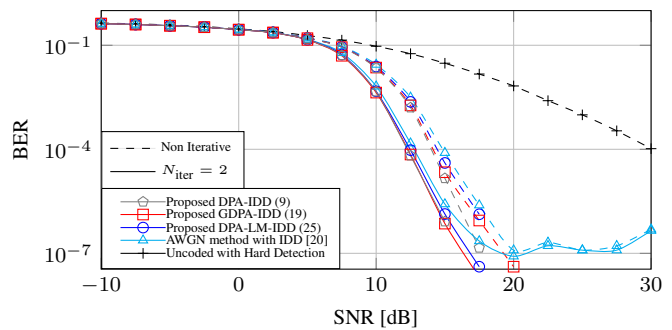}
\caption{Coded BER versus $\mathrm{SNR},\ K=2,\ M=6, \ \alpha_s=8, \ \alpha_x=4$} 
\label{fig:BER_K2_M10}       
\end{figure}

{As can be seen in Fig.~\ref{fig:BER_K2_M10}, all proposed methods provide similar performance for low-SNR. As expected, for the high-SNR regime the proposed DPA-IDD method, that relies on the true PDF of the received signal, yields a lower BER as compared with the proposed suboptimal methods.
Furthermore, considering the marginal performance loss referring to the proposed DPA-IDD method, shown in Fig.~\ref{fig:BER_K2_M10}, reasonable complexity performance trade-offs can be achieved via using the proposed suboptimal methods.}

{The BER performance associated with the system that uses the common AWGN soft detector is similar to the proposed methods for low-SNR. However, in the medium and high-SNR regime, the distortion brought by the discrete precoding becomes relevant, and, since this is not considered in the common AWGN receive processing it results in an error floor in the BER, as shown in Fig.~\ref{fig:BER_K2_M10}.} \looseness-1

{Finally, Fig.~\ref{fig:BER_K2_M10} shows an improvement in performance when using the iterative method. With a relatively small number of iterations there is a gain of approximately $1.5 \ \text{dB}$ when compared with the non iterative approach.}
\section{Conclusions}
\label{sec:conclusions}

{This study proposes three soft detection approaches which calculate extrinsic information values that are used for computing the LLRs via the DPA-IDD scheme.
Numerical results show that employing the common LLR computation method for AWGN channels without taking into account the effects of the discrete precoder causes an error floor in the systems' BER for high-SNR.
By relying on more sophisticated LLR computation methods, the proposed approaches mitigate this problem while also enhancing the overall BER performance of the system.}
\newpage

\bibliographystyle{IEEEbib}
\bibliography{ref}

\end{document}